\documentclass[letterpaper,preprintnumbers,english,floats,floatfix,amssymb,prd,superscriptaddress,twocolumn,nofootinbib]{revtex4}
\usepackage{amssymb,amsmath}
\usepackage{epsfig,hyperref}
\usepackage[svgnames]{xcolor}
\usepackage{pgf,tikz}
\usepackage{epstopdf}
\usepackage[british]{babel}
\usepackage{enumerate}
\usepackage{enumitem}
\usepackage{makecell}
\usepackage{booktabs}
\usepackage{graphicx}
\usepackage{nicefrac}
\usepackage{units}
\usepackage{mathrsfs}
\usepackage{placeins}

\makeatletter
\DeclareRobustCommand*{\bfseries}{%
	\not@math@alphabet\bfseries\mathbf
	\fontseries\bfdefault\selectfont
	\boldmath
}
\makeatother

\def\be{\begin{equation}}
\def\ee{\end{equation}}
\def\beq{\begin{eqnarray}}
\def\eeq{\end{eqnarray}}

\makeatletter

\newcommand{\arXiv}[2][]{\href{http://arxiv.org/abs/#2}{\texttt{arXiv:#2\@ifempty{#1}{}{ [#1]}}}}
\makeatother

\begin{document}
\title{Energy transfer in the collision of two scalar wave packets in spherical symmetry}
	
\author{Li-Jie Xin}
\affiliation{MOE Key Laboratory of Fundamental Physical Quantities Measurement, Hubei Key Laboratory of Gravitation and Quantum Physics, PGMF, and School of Physics, Huazhong University of Science and Technology, Wuhan 430074, Hubei, China}
	
\author{Jun-Qi Guo}%
\email{sps_guojq@ujn.edu.cn}
\affiliation{School of Physics and Technology, University of Jinan, Jinan 250022, Shandong, China}
	
\author{Cheng-Gang Shao\textsuperscript{1}}
\email{cgshao@hust.edu.cn}
	
\date{\today}
	
\begin{abstract}
We study the collisions of two scalar wave packets in the asymptotically flat spacetime and asymptotically anti-de Sitter spacetime in spherical symmetry. An energy transfer formula is obtained, $y=Cm_{i}m_{o}/r$, where $y$ is the transferred energy in the collisions of the two wave packets, $m_i$ and $m_o$ are the Misner-Sharp energies for the ingoing and outgoing wave packets, respectively, $r$ is the areal radius and collision place, and $C=1.873$ and $C=1.875$ for the asymptotically flat spacetime and asymptotically anti-de Sitter spacetime circumstances, respectively. The formula is universal, independent of the initial profiles of the scalar fields.
\end{abstract}
\maketitle

\section{Introduction\label{sec:introduction}}
Gravitational collapse, closely related to the formation of stars, galaxies and large-scale structures, has been an important subject in gravitation and cosmology. During the collapse, the collision of matter fields and energy transfer processes usually occur, significantly affecting the eventual outcomes of collapse.

The dust fluid is a dominant matter field in the Universe. Nakao \emph{et al.}~investigated the collision of two spherical thin shells of dust fluid in the asymptotically flat spacetime, and derived an expression for the energy transfer and change in the 3-momentum~\cite{Nakao:1999qto}. Ida and Nakao extended these results to spherically symmetric spacetime with charge and cosmological constant~\cite{Ida:1999cqw}. Cardoso and Rocha studied the dynamics of two thin shells of perfect fluid confined in a spherical box~\cite{Cardoso:2016wcr} and asymptotically anti-de Sitter (AdS) spacetime~\cite{Brito:2016xvw}, where critical behaviors and chaotic phenomena were displayed.
	
In 2006, Dafermos and Holzegel proposed the conjecture of instability of the AdS spacetime in two talks~\cite{Dafermos_2006,Holzegel_2006}. In 2011, by simulating the evolution of a spherical massless scalar field in the asymptotically AdS spacetime, Bizon and Rostworowski independently obtained the numerical results suggesting that the AdS spacetime is unstable under arbitrarily small generic perturbations~\cite{Bizon_2011}. They observed that the initial parameter could be classified according to the number of round trips in space. When the initial parameter is large enough, the scalar field collapses directly to form a black hole. A scalar field with an arbitrarily small initial amplitude oscillates enough round trips and eventually collapses to form a black hole. Maliborski obtained similar results for the collapse of a massless scalar field confined in a timelike worldtube with a perfectly reflecting wall~\cite{Maliborski_2012}. It was demonstrated that energy tends to gather during the wave packet movement in restricted systems in Refs.~\cite{Bizon_2011,Maliborski_2012,Maliborski:2014rma,Maliborski:2014fbo,Bizon:2015pfa,Jalmuzna:2011qw,Lubbe:2014hpa,Cai:2015jbs,Cai:2016bsa}. Cai \emph{et al.} investigated the gravitational collapse of a massless scalar field and discovered a new power-law behavior for the time of gapped collapse. They studied the cirtical phenomenon near the threshold of black hole formation, in order to better understand the difference between the results in restricted asymptotically flat spacetime and asymptotically AdS spacetime.
	
People investigated the causes to the instability of enclosed spacetimes. In the collapse toward black hole formation, the energy of the matter field will concentrate in a small place. The energy of the matter is transferred from a given wavelength spectrum to short wavelengths. People studied this process using the resonant approximation approach~\cite{Bizon_2011,Balasubramanian:2014cja,Craps:2014vaa,Craps:2014jwa,Bizon:2015pfa}. In this approach, one rewrites the nonlinear dynamical equations with the linearized normal modes. In these equations, there are many rapidly oscillating terms, which cumulative effect can be bounded. However, the remaining terms account for the strong energy transfer between the normal modes whose frequencies satisfy certain resonance relations.

The resonant approximation approach was preliminarily used to explain the instability problem in Ref.~\cite{Bizon_2011}, and was further implemented in Ref.~\cite{Balasubramanian:2014cja}. A fully analytic formulation was derived in Refs.~\cite{Craps:2014vaa,Craps:2014jwa}. In Ref.~\cite{Bizon:2015pfa}, the scalar collapse in five dimensional AdS spacetime was studied, and the resonant approximation reproduces the amplitudes of the true solutions related to the scalar field considerably well.

Moschidis mathematically proved the instability of the AdS spacetime using the Einstein-massless Vlasov system in time domain~\cite{Moschidis_1704,Moschidis_1812}. Moschidis dropped some Vlasov particles into the AdS spacetime, which created concentric waves of matter in spacetime. Among the many concentric waves, the first two waves contain the most matter and energy, so it is sufficient to focus on them. The first wave will expand outward, hit the boundary, and bounce back toward the center. So does the second wave. When the first wave bounces off the boundary and begins to contract toward the origin, it will hit the second wave, which is still expanding. After the first wave arrives at the origin, it will expand again, and cross the second wave that is still contracting. Moschidis showed that the expanding wave always transfers energy to the contracting one, with the transferred energy near the origin being greater than near the outer boundary. Consequently, the second wave obtains more and more energy from the first one. Eventually, a black hole forms. For reviews on the instability of the AdS spacetime, see Refs.~\cite{Martinon:2017ppj,Evnin:2021buq}. In this paper, we simulate the collisions between two scalar wave packets in the asymptotically flat spacetime and asymptotically AdS spacetime, and obtain universal energy transfer formulas.

The paper is organized as follows. In Sec.~\ref{sec:methodology}, we describe the methodology on simulating collisions of two scalar wave packets in the asymptotically flat spacetime and asymptotically AdS spacetime. In Secs.~\ref{sec:flat} and \ref{sec:ads}, we report the energy transfer formulas for the collisions in the two spacetimes, respectively. The results are summarized in Sec.~\ref{sec:summary}. Throughout the paper, we set $4\pi G=c=1$.

\section{Methodology\label{sec:methodology}}
\subsection{Asymptotically flat spacetime}
We first consider the collision of two scalar wave packets in the asymptotically flat spacetime in spherical symmetry with a reflecting wall on the outer boundary. The scalar fields possess a potential $V(\phi)$. The action for the system is described as
\be
S_{1} = \int d^{4}x\sqrt{-g}\left(\frac{R}{16\pi G}-\frac{1}{2}\nabla^{\mu}\phi\nabla_{\mu}\phi-V(\phi)\right). \label{action_flat}
\ee
Einstein's equation and the equation of motion of the scalar field $\phi$ are
\be
R_{\mu\nu}-\frac{1}{2}g_{\mu\nu}R=8\pi G T_{\mu\nu},
\label{Einstein_flat}
\ee
\be
\nabla^{\mu}\nabla_{\mu}\phi = V'(\phi),
\label{Eom_flat}
\ee
where $T_{\mu\nu}$ is the energy-momentum tensor for the scalar field,
\be
T_{\mu\nu} = \nabla_{\mu}\phi\nabla_{\nu}\phi -g_{\mu\nu}\left(V(\phi)+\frac{1}{2}(\nabla\phi)^{2}\right).
\label{Emt_flat}
\ee

We run the simulation in the coordinates~\cite{Bizon_2011},
\be
ds^{2}=-A\left(r,t\right)e^{-2\delta\left(r,t\right)}dt^{2}+\frac{1}{A\left(r,t\right)}dr^{2}+r^{2}d\Omega^{2}.
\label{metric_flat}
\ee
Then the equations are
\begin{flalign}
&&A_{,r}&=\frac{1-A}{r}-rA({\Phi^{2}+\Pi^{2}})+2rV(\phi),&\label{flat_a}\\
&&\delta_{,r}& = -r\left(\Phi^{2}+\Pi^{2}\right),&\label{flat_d}\\
&&\Pi_{,t}& = \frac{\left(r^{2}Ae^{-\delta}\Phi\right)_{,r}}{r^{2}}-e^{\delta}V'(\phi),&\label{flat_pi}\\
&&\Phi_{,t}& = \left(Ae^{-\delta}\Pi\right)_{,r},&\label{flat_phi}
\end{flalign}
where $(_{,r})$ and $(_{,t})$ denote the partial derivatives with respect to $r$ and $t$, respectively, $\Phi\equiv\partial \phi/\partial r$, and $\Pi\equiv A^{-1}e^{\delta}\partial\phi/\partial t$.

Regarding the boundary conditions, for regularity concern, we set $A=1$ and $\Phi=0$ at the center. We use the normalization $\delta=0$ at the center, so that $t$ is the proper time at the center. On the outer boundary, we set $A=\mathrm{Const}$ and $\phi=\mathrm{Const}$ to characterize a nondissipative physical system with a mirror on the outer boundary. The initial conditions are set up as below:
\be
\begin{split}
\mathrm{Outgoing}: \phi_{o}|_{t=0}&=0,\\
\Pi_{o}|_{t=0}&=\epsilon_{1}\exp\left[-\frac{\tan^{2}\frac{\pi}{2}\left(r-r_{1}\right)}{\sigma_{1}^{2}}\right].
\end{split}
\label{outgoingwave}
\ee
\be
\begin{split}
\mathrm{Ingoing}: \phi_{i}|_{t=0}&=\epsilon_{2} \exp\left[-\frac{\tan^{2}\frac{\pi}{2}\left(r-r_{2}\right)}{\sigma_{2}^{2}}\right],\\
\Pi_{i}|_{t=0}&=0.
\end{split}
\label{ingoingwave}
\ee
We change the energy of the outgoing wave and ingoing wave by varying $\epsilon_{1}$, $\sigma_{1}$, $r_{1}$ and $\epsilon_{2}$, $\sigma_{2}$, $r_{2}$, respectively.
We use the Misner-Sharp energy~\cite{Misner_1964} to describe the total energy inside a sphere of radius $r$,
\be
\begin{split}
m&\equiv\frac{r}{2}(1-g^{\mu\nu}r_{,\mu}r_{,\nu})\\
&=\frac{r}{2}(1-A)\\
&=\frac{1}{2}\int^{r}_{0}A(\Phi^{2}+\Pi^{2})r'^{2}dr'.
\end{split}
\label{MS}
\ee

Denote $m_i$ and $m_o$ as the Misner-Sharp energies for the spacetimes occupied by the (outer) ingoing and (inner) outgoing wave packets, respectively. We define the point on which $|\Phi|$ is smaller by a factor of 3 orders of magnitude than the maximum value of $|\Phi|$ in the two wave packets as the separation point between the two packets.
\be m_o\equiv m|_{r=r_o}, \hphantom{ddd} m_i\equiv m|_{r=r_b}-m_o,\ee
where $r_o$ is the separation point and $r_b$ is the outer boundary. The transferred energy is defined as the energy change for the wave packets before and after the collision.

In the simulation, we integrate Eqs.~(\ref{flat_a}) and (\ref{flat_d}) by the fourth-order accurate finite-difference method, and evolve Eqs.~(\ref{flat_pi}) and (\ref{flat_phi}) by the fourth-order Runge-Kutta method. The numerical code we use is described in detail in Ref.~\cite{Maliborski:2014fbo}.

\begin{figure*}[t!]
	\centering
	\begin{tabular}{cc}
		\includegraphics[width=0.8\textwidth]{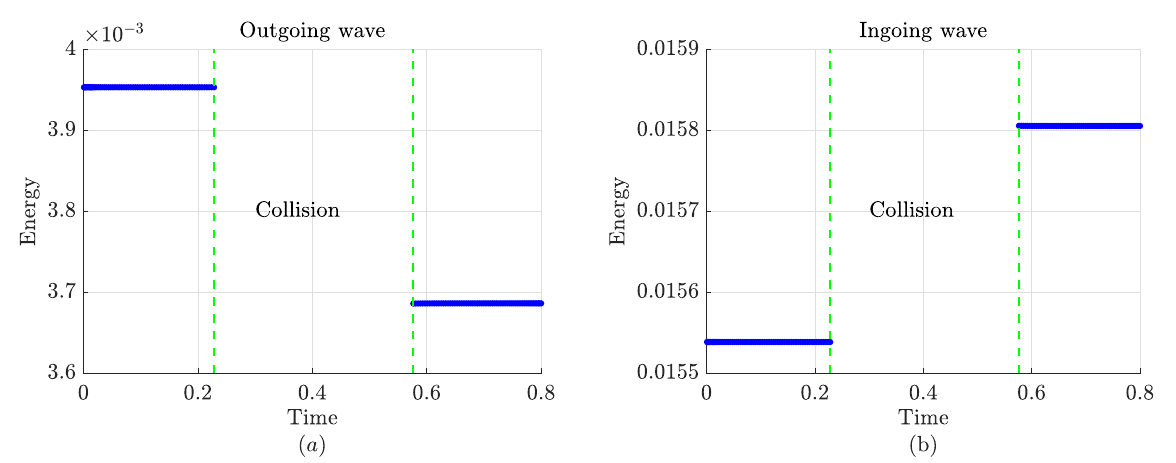}
	\end{tabular}
	\caption{The change of energy for the (a) outgoing and  (b) ingoing wave packets during the collision of two shells of massless scalar fields in the asymptotically flat spacetime. Parameters in the initial profiles of the scalar fields in Eqs.~(\ref{outgoingwave}) and (\ref{ingoingwave}): $\epsilon_{1}=200$, $\sigma_{1}=1/50$, $r_{1}=0$,
		$\epsilon_{2}=\sigma_{2}=1/50$, and $r_{2}=0.9$. The blank area in the middle part corresponds to the collision process.}
	\label{fig:change}
\end{figure*}

\subsection{Asymptotically AdS spacetime}
Now we consider the collision of two massless scalar fields in the asymptotically AdS spacetime in spherical symmetry. We follow the methodology in Ref.~\cite{Bizon_2011}. The action governing the dynamics of the system is expressed as
\be
S_{2}=\int d^{4}x\sqrt{-g}\left(\frac{R-2\Lambda}{16\pi G}-\frac{1}{2}\nabla^{\mu}\phi\nabla_{\mu}\phi\right),
\ee
where $\Lambda$ is the negative cosmological constant. Then the equations of motion for the system are
\be
R_{\mu\nu}-\frac{1}{2}g_{\mu\nu} R +\Lambda g_{\mu\nu}= 8\pi G T_{\mu\nu},
\label{einstein_ads}
\ee
\be
\nabla^{\mu}\nabla_{\mu}\phi = 0.
\label{eom_ads}
\ee
We use the coordinates
\be
\begin{split}
ds^{2}=\frac{l^{2}}{\cos^{2}x}\Big[&-A(x,t)e^{-2\delta(x,t)}dt^{2}+\frac{1}{A(x,t)}dx^{2}\\
& +\sin^{2}xd\Omega^{2}\Big], \hphantom{ddd} x\in\left[0,\hphantom{d} \frac{\pi}{2}\right],
\end{split}
\label{metric_ads}
\ee
where the AdS length scale $l$ is related to $\Lambda$ by $l^{2}=-3/\Lambda$. Then we obtain the equations
\begin{flalign}
&&A_{,x}&=\frac{1+2\sin^{2}x}{\sin x \cos x}\left(1-A\right)-\sin x \cos x A\left(\Phi^{2}+\Pi^{2}\right),\label{ads_a}\\
&&\delta_{,x}&=-\sin x \cos x\left(\Phi^{2}+\Pi^{2}\right),\label{ads_d}\\
&&\Pi_{,t}&=\frac{\left(\tan^{2}xAe^{-\delta}\Phi\right)_{,x}}{\tan^{2}x},\label{ads_pi}\\
&&\Phi_{,t}&=\left(A e^{-\delta}\Pi\right)_{,x},\label{ads_phi}
\end{flalign}
where $\Phi\equiv\partial \phi/\partial x$, and $\Pi\equiv A^{-1}e^{\delta}\partial\phi/\partial t$. It is notable that the length scale $l$ only appears in the definition of energy and is absent in the equations of motion.

Regarding the boundary conditions, for regularity concern, we set $A=1$ and $\Phi=0$ at the center. We use the normalization $\delta=0$ at the center. For regularity concern, we set $A=1$ on the outer boundary $x=\pi/2$. In addition, we set $\phi=\mathrm{Const}$ on the outer boundary. The initial conditions are set up as below:
\be
\begin{split}
\mathrm{Outgoing}: \phi_{o}|_{t=0}&=0,\\
\Pi_{o}|_{t=0}&=\epsilon_{3}\exp\left[-\frac{\tan^{2}\frac{\pi}{2}\left(x-x_{3}\right)}{\sigma_{3}^{2}}\right],
\end{split}
\label{outgoingwave_ads}
\ee
\be
\begin{split}
\mathrm{Ingoing}: \phi_{i}|_{t=0}&=\epsilon_{4}\exp\left[-\frac{\tan^{2}\frac{\pi}{2}\left(x-x_{4}\right)}{\sigma_{4}^{2}}\right],\\
\Pi_{i}|_{t=0}&=0.
\end{split}
\label{ingoingwave_ads}
\ee
The Misner-Sharp energy $m$ within a sphere of radius $r(=l\tan x)$ can be written as~\cite{deOliveira:2012dt}
\be
\begin{split}
	m&\equiv\frac{r}{2}\left(1+\frac{r^{2}}{l^{2}}-g^{\mu\nu}r_{,\mu}r_{,\nu}\right)\\
	&=\frac{l(1-A)\sin x}{2\cos^{3}x}\\
	&=\frac{l}{2}\int^{x}_{0}A(\Phi^{2}+\Pi^{2})\tan^{2}x'dx'.
\end{split}
\label{MS_ads}
\ee

For simplicity, we set $l=1$. In the simulation, we integrate Eqs.~(\ref{ads_a}) and (\ref{ads_d}) by the fourth-order accurate finite-difference method, and evolve Eqs.~(\ref{ads_pi}) and (\ref{ads_phi}) by the fourth-order Runge-Kutta method.

\begin{figure*}[t!]
	\centering
	\begin{tabular}{cc}
		\includegraphics[width=0.95\textwidth]{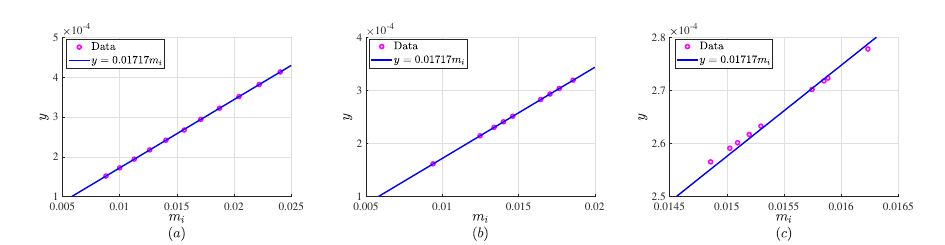}
	\end{tabular}
	\caption{The transferred energy $y$ vs the initial energy $m_i$ of the ingoing wave packet with $V(\phi)=0$ in the asymptotically flat spacetime. We fix the parameters in the initial condition for the outgoing wave~(\ref{outgoingwave}), $\epsilon_{1}=200$, $\sigma_{1}=1/50$, $r_{1}=0$, and vary those for the ingoing one~(\ref{ingoingwave}), $\epsilon_{2}$, $\sigma_{2}$, $r_{2}$, respectively. (a), (b) and (c) Results obtained by varying $\epsilon_{2}$, $\sigma_{2}$ and $r_{2}$, respectively.}
	\label{fig:y_mi}
\end{figure*}

\begin{figure*}[t!]
	\centering
	\begin{tabular}{cc}
		\includegraphics[width=0.95\textwidth]{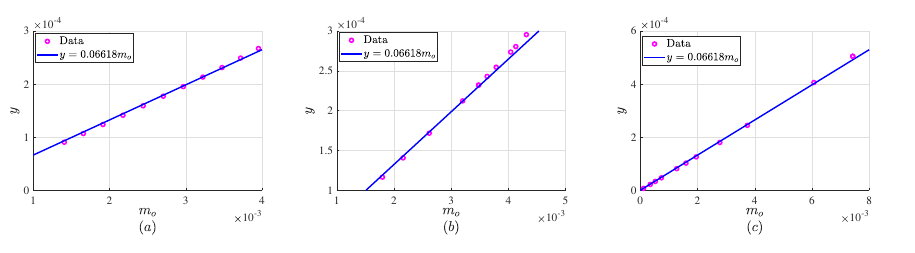}
	\end{tabular}
	\caption{The transferred energy $y$ vs the initial energy $m_o$ of the outgoing wave packet with $V(\phi)=0$ in the asymptotically flat spacetime. We fix the parameters in the initial condition for the ingoing wave~(\ref{ingoingwave}), $\epsilon_{2}=1/500$, $\sigma_{2}=1/50$, $r_{2}=0.9$, and vary those for the outgoing one~(\ref{outgoingwave}), $\epsilon_{1}$, $\sigma_{1}$, $r_{1}$, respectively. (a), (b) and (c) Results obtained by varying $\epsilon_{1}$, $\sigma_{1}$ and $r_{1}$, respectively.}
	\label{fig:y_mo}
\end{figure*}

\section{Result I: energy transfer in the asymptotically flat spacetime\label{sec:flat}}
\subsection{Energy transfer for $V(\phi)=0$}
By numerically integrating Eqs.~(\ref{flat_a})~(\ref{flat_phi}), we simulate the collision of two scalar wave packets in the asymptotically flat spacetime. In this subsection, we set $V(\phi)\equiv0$. We impose the outer boundary at $r=1$. The first wave packet is placed near the origin and will move outwards, and the second one is put near the outer boundary and will move inwards. As shown in Fig.~\ref{fig:change}, we observe that the transmitted energy during the collision of the two packets always flows from the outgoing wave to the ingoing one, which is consistent with the results obtained in Refs.~\cite{Moschidis_1704,Moschidis_1812}.

\begin{figure*}[t!]
	\centering
	\begin{tabular}{cc}
		\includegraphics[width=0.8\textwidth]{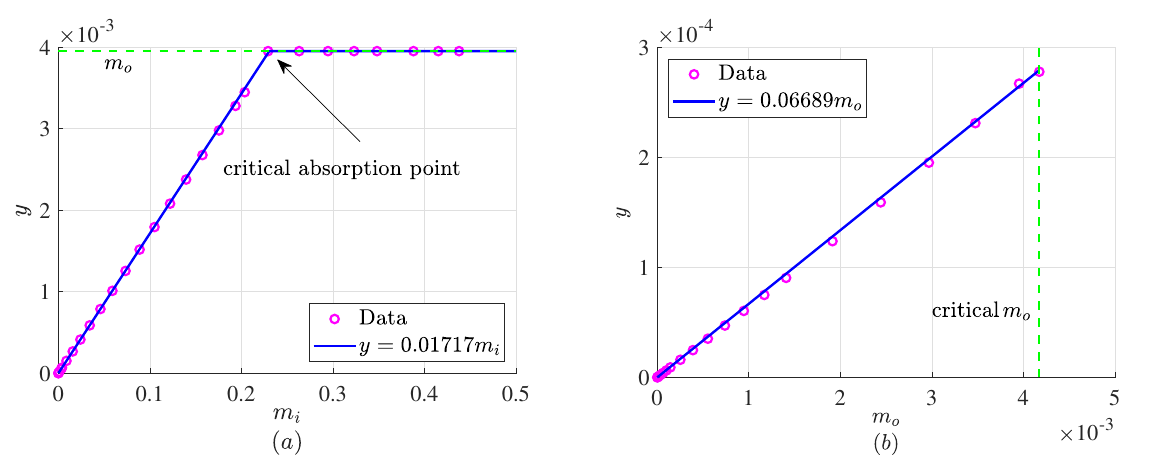}
	\end{tabular}
	\caption{The results of (a) $y$ vs $m_i$ and (b) $y$ vs $m_o$ when the ranges of $m_i$ and $m_o$ are expanded in comparison with Figs.~\ref{fig:y_mi} and \ref{fig:y_mo}.
  (a) We fix the parameters in the initial condition for the outgoing wave~(\ref{outgoingwave}), $\epsilon_{1}=200$, $\sigma_{1}=1/50$, $r_{1}=0$, and vary $\epsilon_{2}$ for the ingoing one~(\ref{ingoingwave}). The critical absorption point signifies the critical collision of the two waves, which just leads to black hole formation. The horizontal line indicates that the (outer) ingoing wave collapses to form a black hole before collisions, preventing the escape of the (inner) outgoing wave from the black hole.
  (b) We fix the parameters in the initial condition for the ingoing wave~(\ref{ingoingwave}), $\epsilon_{2}=1/500$, $\sigma_{2}=1/50$, $r_{2}=0.9$, and vary $\epsilon_{1}$ for the outgoing one~(\ref{outgoingwave}). When $m_o$ is greater than the critical value, the (inner) outgoing wave will directly form a black hole before colliding with the (outer) ingoing one.}
	\label{fig:bdy}
\end{figure*}

\begin{table*}
	\centering
	\begin{tabular}{c|c|c}	
		\hline
		Initial conditions & Outgoing wave & Ingoing wave \\
		\hline
		\rule{0pt}{2em}	Set 1 & $\phi_{o}|_{t=0} = 0$, \hphantom{d} $\Pi_{o}|_{t=0}=\epsilon_{1}\exp\left[-\frac{\tan^{2}\frac{\pi}{2}\left(r-r_{1}\right)}{\sigma_{1}^{2}}\right]$ & $\phi_{i} |_{t=0} = \epsilon_{2} \exp\left[-\frac{\tan^{2}\frac{\pi}{2}\left(r-r_{2}\right)}{\sigma_{2}^{2}}\right]$, \hphantom{d} $\Pi_{i}|_{t=0} = 0$ \\
		
		\hline
		\rule{0pt}{2em}
		Set 2 & $\phi_{o}|_{t=0} = 0$, \hphantom{d} $\Pi_{o}|_{t=0} = \epsilon_{1} r^{3} \exp\left[-\frac{\left(r-r_{1}\right)^{2}}{\sigma_{1}^{2}}\right]$ & $\phi_{i} |_{t=0} = \epsilon_{2} r^{3} \exp\left[-\frac{\left(r-r_{2}\right)^{2}}{\sigma_{2}^{2}}\right]$, \hphantom{d} $\Pi_{i}|_{t=0} = 0$ \\
		
		\hline
		\rule{0pt}{2em}
		Set 3 & $\phi_{o}|_{t=0}=\epsilon_{1}\tanh\left(\frac{r_{1}-r}{\sigma_{1}}\right)$, \hphantom{d} $\Pi_{o}|_{t=0} = 0$ & $\phi_{i} |_{t=0} = 0$, \hphantom{d} $\Pi_{i}|_{t=0} = \epsilon_{2} \exp\left[-\frac{\tan^{2}\left(r-r_{2}\right)}{\sigma_{2}^{2}}\right]$ \\
		
		\hline
	\end{tabular}
	\caption{Three sets of initial conditions for the outgoing~(\ref{outgoingwave}) and ingoing~(\ref{ingoingwave}) wave packets.}
	\label{tab:initial}
\end{table*}

We fix the parameters in the initial condition for the outgoing wave~(\ref{outgoingwave}), and vary those for the ingoing packet~(\ref{ingoingwave}). As shown in Fig.~\ref{fig:y_mi}, we find that the transferred energy $y$ during the collisions has a simple expression with respect to the mass of the ingoing wave packet $m_i$,
\be y=(0.01717\pm0.00002) m_{i}.\label{y_mi}\ee
Similarly, as shown in Fig.~\ref{fig:y_mo}, by fixing the parameters in the initial condition for the ingoing wave~(\ref{ingoingwave}), and varying those for the outgoing one~(\ref{outgoingwave}), we have
\be y=(0.06689\pm0.00068) m_{o}.\label{y_mo}\ee

In Figs.~\ref{fig:y_mi} and \ref{fig:y_mo}, the ranges for the masses, $m_i$ and $m_o$, are very limited. Actually, as shown in Fig.~\ref{fig:bdy}, when we expand the ranges for $m_i$ and $m_o$ till black hole formation, the results~(\ref{y_mi}) and (\ref{y_mo}) remain valid. Some details are the following:
\begin{enumerate}[fullwidth,itemindent=0em,label=(\roman*)]
  \item When the initial mass of the (outer) ingoing wave is large enough, the ingoing wave will collapse to form a black hole directly before colliding with the (inner) outgoing wave. In this circumstance, the outgoing wave will be totally absorbed into the black hole. So the transferred energy $y$ is equal to the initial mass of the outgoing wave $m_o$, which generates the horizontal line in Fig.~\ref{fig:bdy}(a). In Fig.~\ref{fig:bdy}(a), the critical absorption point signifies the critical collision of the two waves, which just leads to black hole formation.
  \item We also increase the initial mass of the (inner) outgoing wave $m_o$. Certainly, when $m_o$ is large enough, the outgoing wave will directly form a black hole before colliding with the (outer) ingoing wave. See Fig.~\ref{fig:bdy}(b).
\end{enumerate}

\begin{figure}[t!]
	\centering
	\begin{tabular}{cc}
		\includegraphics[width=0.4\textwidth]{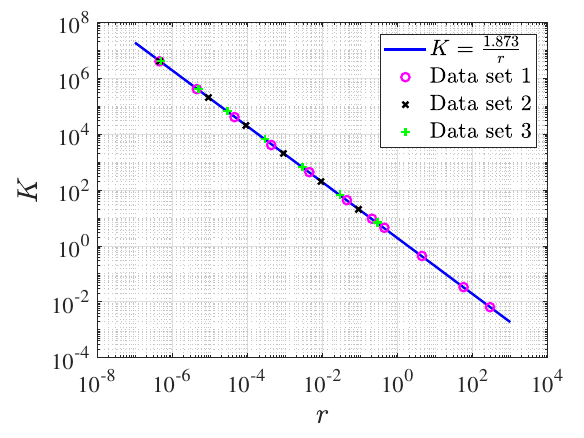}
	\end{tabular}
	\caption{The function $K$ in Eq.~(\ref{y_K}) vs $r$. $K=1.873/r$.  The result is universal, independent of the initial profiles of the wave packets described in Table~\ref{tab:initial}.}
	\label{fig:K}
\end{figure}

The combination of Eqs.~(\ref{y_mi}) and (\ref{y_mo}) generates
\be y=K(r)m_{i}m_{o}.\label{y_K}\ee
By fitting the numerical results of $y/(m_{i}m_{o})$ vs $r$, as shown in Fig.~\ref{fig:K}, we obtain
\be y=(1.873\pm0.050)\frac{m_{i}m_{o}}{r}.\label{flat_final}\ee
We run the simulation with three sets of initial data for the two wave packets as described in Table~\ref{tab:initial}. As shown in Fig.~\ref{fig:K}, the results remain the
same. So the formula~(\ref{flat_final}) is universal and independent of the initial profiles of the wave packets.

In Ref.~\cite{Nakao:1999qto}, Nakao \emph{et al.} studied the collision of two spherical thin shells of dust. An energy transfer formula was derived,
\be y=-u^{a}_{i}u_{oa}\frac{m_{i}m_{o}}{r},\label{y_dust}\ee
where $u^{a}_{i}$ and $u^{a}_{o}$ are the 4-velocities for the shells and other quantities are defined in the same way as above. With Eqs.~(\ref{flat_final}) and (\ref{y_dust}), one can see that the energy transfer formulas for the collisions of scalar field and dust are very close.

\subsection{Energy transfer for $V\left(\phi\right)=\frac{1}{2}\mu^{2}\phi^{2}$}
\begin{figure*}[t!]
	\centering
	\begin{tabular}{cc}
		\includegraphics[width=0.95\textwidth]{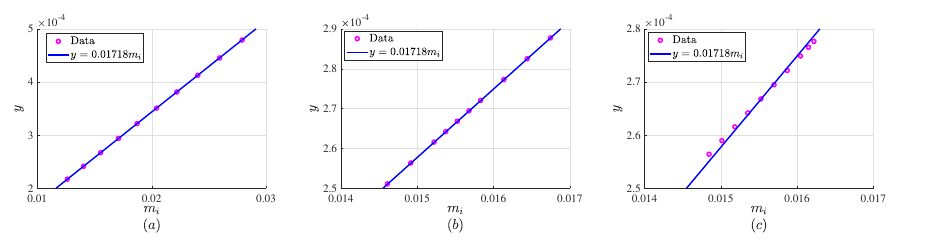}
	\end{tabular}
	\caption{The transferred energy $y$ vs the initial energy $m_i$ of the ingoing wave packet with $V(\phi)=(1/2)\mu^2\phi^2$ in the asymptotically flat spacetime. $\mu^2=5$. We fix the parameters in the initial condition for the outgoing wave~(\ref{outgoingwave}), $\epsilon_{1}=200$, $\sigma_{1}=1/50$, $r_{1}=0$, and vary those for the ingoing one~(\ref{ingoingwave}), $\epsilon_{2}$, $\sigma_{2}$, $x_{2}$, respectively. (a), (b) and (c) Results obtained by varying $\epsilon_{2}$, $\sigma_{2}$ and $r_{2}$, respectively.}
	\label{fig:outgoing_mass}
\end{figure*}

\begin{figure}[t!]
	\centering
	\begin{tabular}{cc}
		\includegraphics[width=0.4\textwidth]{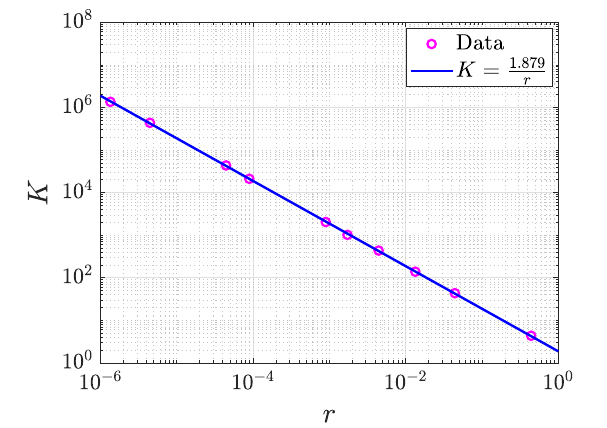}
	\end{tabular}
	\caption{The function $K$ in Eq.~(\ref{y_Km}) vs $r$. $V\left(\phi\right)=(1/2)\mu^{2}\phi^{2}$ and $\mu^{2} = 5$. $K=1.879/r$.}
	\label{fig:K_massive}
\end{figure}

We also explore the collision of two massive scalar wave packets. Let the potential $V(\phi)$ in Eq.~(\ref{action_flat}) take the form
\be V\left(\phi\right)=\frac{1}{2}\mu^{2}\phi^{2},\ee
and set $\mu^{2}=5$. We put the two wave packets in the same places as in the massless scalar field circumstance. We obtain
\be y=(0.01718\pm0.00004)m_{i},\label{y_mi_mass}\ee
\be y=(0.06614\pm0.00068)m_{o},\label{y_mo_mass}\ee
\be y=K(r)m_{i}m_{o}=(1.879\pm0.042)\frac{m_{i}m_{o}}{r}.\label{y_Km}\ee
See Figs.~\ref{fig:outgoing_mass} and \ref{fig:K_massive}. The coefficients in Eqs.~(\ref{y_mi_mass})~(\ref{y_Km}) are very close to those in the massless scalar field circumstance (\ref{y_mi})~(\ref{flat_final}). We also vary the value of $\mu^{2}$ from 1 to 15, and get similar results as the case of $\mu^{2}=5$. See Table~\ref{tab:mass}.

\begin{figure*}[t!]
	\centering
	\begin{tabular}{cc}
		\includegraphics[width=0.95\textwidth]{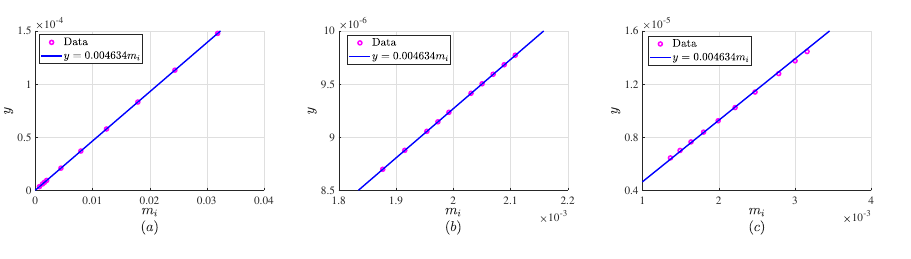}
	\end{tabular}
	\caption{The transferred energy $y$ vs the initial energy of the ingoing wave packet $m_i$ in the asymptotically AdS spacetime. We fix the parameters in the initial condition for the outgoing wave~(\ref{outgoingwave_ads}), $\epsilon_{3}=200$, $\sigma_{3}=1/50$, $x_{3}=0$, and vary those for the ingoing one~(\ref{ingoingwave_ads}), $\epsilon_{4}$, $\sigma_{4}$, $x_{4}$, respectively. (a), (b) and (c) Results obtained by varying $\epsilon_{4}$, $\sigma_{4}$ and $x_{4}$, respectively.}
	\label{fig:y_mi_ads}
\end{figure*}

\begin{figure*}[t!]
	\centering
	\begin{tabular}{cc}
		\includegraphics[width=0.95\textwidth]{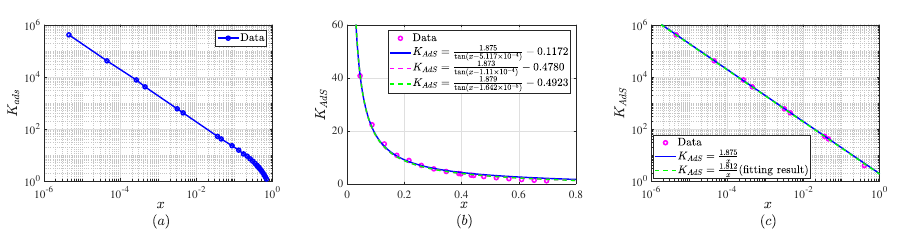}
	\end{tabular}
	\caption{The function $K_{AdS}$ in Eq.~(\ref{ads_pro}) vs. $x$. (a) Numerical data for $K_{AdS}$ vs. $x$. (b) The fitting result for $K_{AdS}$ vs. $x$ at large $x$. (c) The fitting result for $K_{AdS}$ vs. $x$ at small $x$.}
	\label{fig:K_ads}
\end{figure*}

\begin{table*}
	\centering
	\begin{tabular}{c|c|c|c}	
		\hline
		$\mu^{2}$ & Coefficient of $y$ $\propto$ $m_{i}$ & Coefficient of $y$ $\propto$ $m_{o}$ & Coefficient of $y$ $\propto$ $m_{i}m_{o}/r$\\
		\hline
		\rule{0pt}{2em}
		1 & $0.01718\pm0.00004$ & $0.06617\pm0.00068$ & $1.878\pm0.042$ \\
		
		\hline
		\rule{0pt}{2em}
		2 & $0.01718\pm0.00004$ & $0.06616\pm0.00068$ & $1.878\pm0.042$\\
		
		\hline
		\rule{0pt}{2em}
		3 & $0.01718\pm0.00004$ & $0.06616\pm0.00068$ & $1.878\pm0.042$\\
		
		\hline
		\rule{0pt}{2em}
		4 & $0.01718\pm0.00004$ & $0.06615\pm0.00068$ & $1.878\pm0.042$\\
		
		\hline
		\rule{0pt}{2em}
		5 & $0.01718\pm0.00004$ & $0.06614\pm0.00068$ & $1.879\pm0.042$\\
		
		\hline
		\rule{0pt}{2em}
		6 & $0.01718\pm0.00004$ & $0.06612\pm0.00068$ & $1.878\pm0.042$\\
		
		\hline
		\rule{0pt}{2em}
		7 & $0.01717\pm0.00003$ & $0.06610\pm0.00068$ & $1.878\pm0.042$\\
		
		\hline
		\rule{0pt}{2em}
		8 & $0.01717\pm0.00003$ & $0.06606\pm0.00068$ & $1.878\pm0.042$\\
		
		\hline
		\rule{0pt}{2em}
		9 & $0.01716\pm0.00004$ & $0.06601\pm0.00069$ & $1.878\pm0.042$\\
		
		\hline
		\rule{0pt}{2em}
		10 & $0.01715\pm0.00004$ & $0.06654\pm0.00106$ & $1.879\pm0.042$\\
		
		\hline
		\rule{0pt}{2em}
		11 & $0.01713\pm0.00003$ & $0.06585\pm0.00070$ & $1.878\pm0.042$\\
		
		\hline
		\rule{0pt}{2em}
		12 & $0.01711\pm0.00003$ & $0.06574\pm0.00071$ & $1.876\pm0.042$\\
		
		\hline
		\rule{0pt}{2em}
		13 & $0.01707\pm0.00003$ & $0.06558\pm0.00072$ & $1.876\pm0.042$\\
		
		\hline
		\rule{0pt}{2em}
		14 & $0.01703\pm0.00002$ & $0.06539\pm0.00073$ & $1.876\pm0.042$\\
		
		\hline
		\rule{0pt}{2em}
		15 & $0.01698\pm0.00001$ & $0.06515\pm0.00075$ & $1.876\pm0.041$\\
		
		\hline
	\end{tabular}
	\caption{The coefficient $C$ in the energy transfer formula $y=C{m_i}{m_o}/r$ vs $\mu^{2}$ in the function $V(\phi)=(1/2)\mu^2\phi^2$.}
	\label{tab:mass}
\end{table*}

\section{Result II: energy transfer in the asymptotically AdS spacetime\label{sec:ads}}
With the same method implemented above, by integrating Eqs.~(\ref{ads_a})~(\ref{ads_phi}), we obtain that in the asymptotically AdS spacetime circumstance, the transferred energy is also proportional to $m_i$ and $m_o$,
\be y=(0.004634\pm0.000025)m_{i},\ee
\be y=(0.02040\pm0.00044)m_{o}.\ee
As an example, we plot the result of $y\propto m_i$ in Fig.~\ref{fig:y_mi_ads}. Then we obtain
\be y=K_{AdS}(x) m_{i}m_{o}.\label{ads_pro}\ee

Equation~(\ref{MS_ads}) shows that the energies, $m_o$ and $m_i$, are proportional to $l$. On the other hand, the numerical results generate that $y$ is proportional to $m_{i}m_{o}$. So a compatible expression is $y\propto m_{i}m_{o}/l$, which has been verified by the numerical results. As shown in Fig.~\ref{fig:K_ads}(b), by fitting the numerical results of $y/(m_{i}m_{o})$ vs $x$, we obtain
\be y=(1.875\pm0.135)\frac{m_{i}m_{o}}{l\tan x}=(1.875\pm0.135)\frac{m_{i}m_{o}}{r}.\label{ads_K}\ee

Regarding the result~(\ref{ads_K}), we make the following discussions:
\begin{enumerate}[fullwidth,itemindent=0em,label=(\roman*)]
  \item We examine the validity of the result~(\ref{ads_K}). At the limit $x\rightarrow0$, Eq.~(\ref{ads_K}) is reduced to
\be y\approx 1.875\frac{m_{i}m_{o}}{lx}\approx 1.875\frac{m_{i}m_{o}}{r},\label{y_ads_small_x}\ee
which is very close to the fitting results shown in Fig.~\ref{fig:K_ads}(c),
\be y=(1.812\pm0.099)\frac{m_{i}m_{o}}{lx}\approx 1.812\frac{m_{i}m_{o}}{r}.\label{y_ads_small_x_fit}\ee
Note that at the limit $x\rightarrow0$, the metric for the asymptotically AdS spacetime~(\ref{metric_ads}) is reduced to the form for the asymptotic flat spacetime~(\ref{metric_flat}). Correspondingly, Eqs.~(\ref{y_ads_small_x}) and (\ref{y_ads_small_x_fit}) are very close to the energy transfer formula in the asymptotically flat spacetime (\ref{flat_final}).

According to Eq.~(\ref{ads_K}), the transmitted energy asymptotes to zero at the limit $x\rightarrow \pi/2$. This matches with the fact that $x=\pi/2$ corresponds to the spatial infinity of the AdS spacetime.

\item In Refs.~\cite{Moschidis_1704,Moschidis_1812}, Moschidis mathematically proved that in the collision of two shells of Vlasov particles in the AdS spacetime, the transferred energy near the origin is greater than that near the outer boundary. Equation~(\ref{ads_K}) matches well with Moschidis' results.

\item Based on Eqs.~(\ref{flat_final}) and (\ref{ads_K}), we speculate that the energy transfer formula, $y=Cm_{i}m_{o}/r$, is valid in general spherically symmetric spacetimes.
\end{enumerate}

\section{Summary\label{sec:summary}}
The energy transfer during the collisions of matter fields is a basic physical process in gravitational collapse, and affects significantly the final outcomes of gravitational collapse.

With numerical simulations, we studied the collisions of two scalar wave packets in the asymptotically flat spacetime and asymptotically AdS spacetime. By fitting the numerical results, we obtained succinct and universal formulas for the energy transfer during the collisions of matter fields. Such expressions bring us further information on gravitational collapse and help to interpret the instability of the enclosed spacetimes. We speculate that in general spherically symmetric spacetimes, the energy transfer can be written in a unified format, $y=Cm_{i}m_{o}/r$.

\section*{ACKNOWLEDGMENTS}\small
The authors are very thankful to Oleg Evnin and Lin Zhang for the helpful discussions. L.J.X. and C.G.S. are supported by the National Natural Science Foundation of China (Grant No. 11925503). J.Q.G. is supported by Shandong Province Natural Science Foundation under Grant No. ZR2019MA068.

\FloatBarrier	

\end{document}